# Anomalous Critical Behavior of Driven Disordered Systems Beyond the Overdamped Limit

Giuseppe Petrillo,[1] Eduardo Jagla,[2] Eugenio Lippiello,[3] and Alberto Rosso[4]

[1]*Earth Observatory of Singapore (EOS), Nanyang Technological University, Singapore*
[2]*Centro Atómico Bariloche, Instituto Balseiro, Comisión Nacional de Energía Atómica, CNEA, CONICET, UNCUYO, Av. E. Bustillo 9500 (8400) Bariloche, Argentina*
[3]*Department of Mathematics and Physics, University of Campania "L. Vanvitelli", Caserta, Italy*
[4]*LPTMS, CNRS, Université Paris-Saclay, Orsay, 91405, France*

We investigate the role of relaxation mechanisms in the driven response of elastic disordered interfaces in finite dimensions, focusing on the interplay between dimensionality and interaction range. Through extensive numerical simulations, we identify two distinct dynamical regimes. In two-dimensional systems with long-range interactions, we observe a regime of coexistence between pinned and flowing states. In contrast, for one-dimensional interfaces with long-range elasticity, as well as for short-range interactions in both 1D and 2D, the coexistence regime is absent. Nevertheless, the avalanche statistics differ significantly from those of overdamped systems: the usual power-law distribution is replaced by a pronounced bump, associated with large, anomalous avalanches that expand ballistically. We interpret these events as failed synchronization attempts and suggest they could be detected in experimental systems.

The mechanical response of disordered materials to external driving forces is highly intermittent, dominated by sudden collective rearrangements known as *avalanches*. These abrupt events are observed across a wide range of systems, from domain wall motion [1, 2] and crack propagation [3, 4] to the deformation of amorphous solids [5] and fault dynamics [6, 7]. This behavior originates from the complex energy landscape of these materials, which is characterized by a multitude of metastable configurations. When an external force is applied, the energy landscape tilts, progressively destabilizing local minima. As some metastable states disappear, the system undergoes abrupt rearrangements, leading to avalanches that span a wide range of sizes.

In the overdamped limit, the dynamics is governed by gradient descent, ensuring that the system relaxes towards the nearest local minimum. For elastic interfaces driven by an external force $F$, a critical threshold $F_c$ separates two distinct regimes [8–10]. For $F < F_c$, the interface remains pinned in metastable configurations, and avalanches can be triggered. These avalanches exhibit a scale-free size distribution, whose cut-off diverges as $F_c$ is approached. For $F > F_c$, the interface becomes *depinned*, as all metastable states are eliminated, and moves with a finite velocity. A similar behavior is observed in soft amorphous materials under shear: a yield stress $\sigma_y$ must be reached to induce a steady flow in the material [11].

Despite extensive studies, it remains unclear whether the depinning or yielding transitions persist beyond the overdamped limit. Several dynamical mechanisms may come into play in this regime, such as inertia [12? –15], stress overshoot [16–18] viscoelasticity [19–22], or internal relaxation processes [23–26]. In such cases, the temporal evolution is no longer governed solely by the gradient of the energy landscape, but also depends on the dynamical state of the system, which may allow it to overcome local energy barriers. We refer to any of these cases in general as having a *non-overdamped dynamics*.

Two fundamental questions remain unresolved: (i) Can a new dynamical regime emerge, distinct from both steady flow and avalanche-dominated motion? (ii) How do the properties of avalanches qualitatively change when moving beyond the overdamped approximation?

A definitive answer is known only for the fully connected model. In this case, the introduction of dynamical effects leads to a novel dynamical regime, intervening between the flowing and pinned regimes, where there is coexistence between flowing and pinned states: force-controlled driving produces hysteresis[18, 20, 21, 25], while displacement-controlled protocols give rise to stick-slip dynamics, with instabilities that, at variance with normal avalanches, span a finite fraction of the whole system [22, 27, 28].

In this work, we demonstrate that dimensionality and interaction range play a crucial role in determining the dynamical behavior of systems with non-overdamped dynamics. Through extensive numerical simulations, we uncover two distinct scenarios in finite dimensions. For two-dimensional interfaces with long-range interactions, we find a regime of coexistence between flowing and pinned states. In addition, we also find important modifications in the avalanche regime, which is now characterized by new critical exponents.

For systems with short-range elasticity (both in 1D and 2D), and for one-dimensional systems with long-range elasticity, the coexistence regime is not observed. Nonetheless, the avalanche dynamics differ markedly from their overdamped counterparts. Notably, the power-law behavior typically seen in avalanche size distributions is replaced by a pronounced bump, associated with large anomalous avalanches that expand ballistically rather than exhibiting erratic dynamics. We interpret these events as "failed attempts" at synchronization [29] and propose that they should be observable in experimental settings.

Our study focuses on the controlled-displacement protocol, with the aim of highlighting the presence or absence of stick-slip dynamics. The case of long-range interactions is discussed in the main text, while the short-range case is addressed in Section II of the Supplementary Material. In the conclusion, we discuss the potential implications of the force-controlled protocol and of non-overdamped dynamics for the shear response of amorphous solids.

*The model.* We consider a $D$-dimensional interface composed of $L^D$ blocks arranged on a periodic lattice (square for $D = 2$). Each block $i$ has a displacement $h_i$, and is subject to a local force $\sigma_i$, resulting from both an external drive and elastic interactions with neighboring blocks:

$$\sigma_i = k_0(w - h_i) + \sum_j G_{ij}(h_j - h_i) \tag{1}$$

The first term is a driving force that pulls each block toward the imposed displacement $w$, with coupling strength $k_0$. The second term accounts for the interaction with other blocks, weighted by the elastic kernel $G_{ij}$.

A block remains pinned as long as the local force $\sigma_i$ is below a random threshold $\sigma_i^{\text{th}}$. We define the global force acting on the interface as the spatial average of local forces:

$$F(w) = \frac{1}{L^D} \sum_i \sigma_i = k_0(w - h_{\text{CM}}) \tag{2}$$

The second equality follows from the fact that elastic forces balance out on average. Here, $h_{\text{CM}}$ is the center of mass of the interface. The system dynamics is characterized by three steps:

*Drive:* During this stage, all blocks are stable. The external load $w$ is gradually increased until a first block, which we denote by index $i$, becomes unstable. This block acts as the hypocenter of the avalanche.

*Avalanche:* Once instability is triggered, the load $w$ is kept constant. The unstable block $i$ slips to the next pinning site ($h_i \to h_i + 1$) and its threshold $\sigma_i^{\text{th}}$ is renewed. The slip redistributes the forces in the system according to:

$$\sigma_i \to \sigma_i - \sum_{j \neq i} G_{ij} - k_0, \tag{3}$$

$$\sigma_j \to \sigma_j + G_{ij}, \quad \forall j \neq i. \tag{4}$$

Following the redistribution in Eq. (4), additional blocks $j \neq i$ may become unstable if their local forces exceeds their threshold, i.e., if $\sigma_j > \sigma_j^{\text{th}}$. These blocks also slip ($h_j \to h_j + 1$), each triggering a new redistribution of the forces according to Eqs. (3) and (4).

The avalanche unfolds over successive *generations* $l = 0, 1, \ldots$, with $n(l)$ denoting the number of blocks slipping at generation $l$. The initial hypocenter block $i$ belongs to generation $l = 0$; blocks destabilized directly by its slip belong to generation $l = 1$, and so on.

The avalanche stops when reaching a generation $l^*$ for which $n(l^*) = 0$, meaning that no further slips occur. In the overdamped limit, the system immediately returns to the drive phase. In the non-overdamped regime, an intermediate *relaxation phase* may take place before a new avalanche is initiated.

*Relaxation:* After an instability has occurred and the local forces have been updated according to Eqs.(3-4), the system undergoes a relaxation phase during which the external load $w$ remains constant, while the local forces evolve dynamically over time. Specifically, considering a time interval $\tau$ after the slip of site $i$, the stress updates as:

$$\sigma_i \to \sigma_i + \Theta \cdot (1 - \Phi(\tau)) \cdot \sum_{j \neq i} G_{ij}, \tag{5}$$

$$\sigma_j \to \sigma_j - \Theta \cdot (1 - \Phi(\tau)) \cdot G_{ij}, \quad \forall j \neq i \tag{6}$$

Here, $\Phi(\tau)$ is a monotonically decreasing function of the elapsed time $\tau$ since the slip, with $\Phi(0) = 1$ and $\Phi(\infty) = 0$.

The parameter $\Theta \in [0, 1]$ controls the strength of the relaxation. From Eqs. (4) and (6), we see that an interacting site $j \neq i$ first experiences a kick $G_{ij}$ in the local force, before losing a fraction $\Theta$ of it through relaxation. The initial kick can destabilize the site $j$ ($\sigma_j > \sigma_j^{th}$), thereby promoting the propagation of a larger avalanche. Conversely, from Eq. (3) we see that the site $i$ that has already slipped is left with a very low local force. Subsequently, according to Eq.(5), $\sigma_i$ increases over time, potentially leading to its reactivation ($\sigma_i > \sigma_i^{th}$) and to the triggering of further failures.

The specific form of $\Phi(\tau)$ determines the nature of the relaxation dynamics. In this work, we consider two limiting cases:

- *Fast dissipation* [16, 18]: Used to model inertial effects, this regime identifies the internal time $\tau$ with the generation index $l$. Specifically, $\Phi(\tau) = 1$ at the generation when block $i$ slips, remains 1 at the next generation, and drops to 0 two generations later. Relaxation thus occurs concurrently with avalanche propagation. Once the avalanche ends, the drive phase resumes immediately.

- *Slow dissipation* [22, 23, 30, 31]: Used to model rate-and-state friction [32–35] or viscoelastic effects, this regime separates relaxation from avalanche propagation. Here, $\Phi(\tau)$ remains constant and equal to 1 throughout the avalanche.

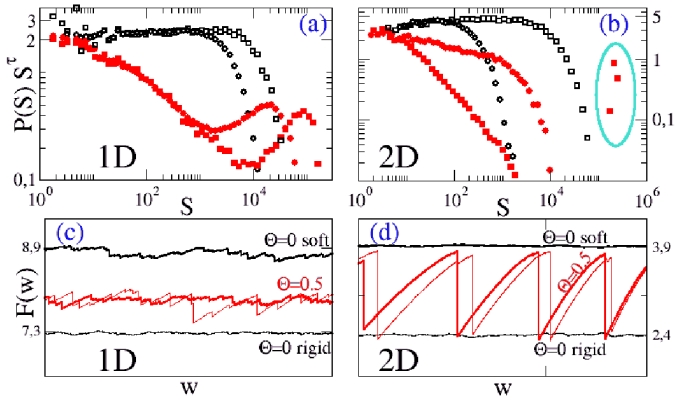

FIG. 1. Distributions $P(S)$ of avalanche sizes (panels a and b) and functions $F(w)$ versus $w$ (panels c and d) for long-range elasticity in 1D (panels a and c) and 2D (panel b and d). Black open symbols correspond to depinning, red filled symbols to the relaxed model with $\Theta = 0.5$.
(a) 1D case with system size $L = 1024$. Circles represent $k_0 = 0.01$, squares $k_0 = 0.005$. (b) 2D case with $L = 200$ and $\Theta = 0.5$. Circles represent $k_0 = 0.2 > k_c$, squares $k_0 = 0.1 < k_c$. The turquoise ellipse highlights system-size avalanches. (c) $F(w)$ in 1D for $k_0 = 0.01$ and $\Theta = 0.5$, with thin continuous lines for $L = 1024$, and a thicker line for $L = 8192$. In black, the depinning models for the rigid (lower curve) and soft (upper curve) interface. (d) $F(w)$ in 2D for $k_0 = 0.1$ and $\Theta = 0.5$, with thin continuous lines for $L = 100$, and a thicker line for $L = 200$. In black, the depinning models for the rigid (lower curve) and soft (upper curve) interface.

Only after the avalanche stops ($n(l^*) = 0$) does $\Phi(\tau)$ start to decrease, generating stress increases that can trigger new failures, called *aftershocks*. After all aftershocks have occurred, the system transitions back to the drive phase.

In this work, we focus primarily on the slow dissipation case while also drawing insights from the fast case. For long-range elasticity, the strength $G_{ij}$ decays with the distance $r$ between two blocks as $1/r^{D+\alpha}$, with $0 < \alpha < 2$. We consider periodic boundary conditions, and the distance $r$ is defined as the shortest distance between two points on the torus. We focus on the physical case $\alpha = 1$: in 1D, this is relevant for propagating crack fronts [36–38], and in 2D, for frictional interfaces [39]. In the Supplementary Material, we study the case $\alpha = 2$ - which corresponds to the short-range limit — and show that it is qualitatively similar to the case 1D for $\alpha = 1$.

*Results–* The nature of the different dynamical regimes is revealed by the statistical properties of the avalanches triggered by increments in $w$. We define the size of the avalanche $S$ as the total number of instabilities recorded between two consecutive increments in the drive. In the presence of slow relaxation, $S$ includes both the main shock and the induced aftershocks; however, considering only the main shocks does not affect any of our conclusions.

We begin by presenting well-known results for $\Theta = 0$, which we refer to as the 'depinning model' throughout the following discussion. These results serve as a baseline to better understand the role of relaxation in the interface dynamics. In Fig. 1a,b, we show the avalanche size distribution $P(S)$, multiplied by $S^\tau$, using the known depinning exponents $\tau = 1.28$ and $\tau = 1.5$ for $D = 1$ and $D = 2$, respectively (see Table I). The approximately flat behavior of $P(S)S^\tau$ confirms the expected power-law decay of $P(S)$ up to a maximum avalanche size $S_{\max}$, beyond which, a faster, exponential-like decay is observed. The upper cutoff $S_{\max}$ depends on the stiffness $k_0$ of the driving spring. To quantify this dependence, in Fig. 2 we report the variation of the avalanche size cutoff $S_{\max}$ as a function of $k_0$. We define the avalanche cutoff as $S_{\max} \equiv \langle S^2 \rangle / \langle S \rangle$, following standard practice. Our results show that $S_{\max}$ diverges in the limit $k_0 \to 0$ according to a power law:

$$S_{\max} \sim k_0^{-\sigma},$$

with the values of the exponent $\sigma$ reported in Table I.

The presence of an upper cutoff $S_{\max}$ affects the global force $F(w)$. According to Eq. (2), each avalanche causes a drop in the external force $F(w)$ by an amount $S/L^D$, while each driving step increases the force by $k_0 \delta w$. These opposing effects balance on average, leading to:

$$\langle \delta w \rangle = \frac{\langle S \rangle}{k_0 L^D}. \tag{7}$$

Accordingly, $\langle \delta w \rangle$ controls the amplitude of force fluctuations. If the average avalanche size $\langle S \rangle$ remains independent of the system size $L$, the fluctuations in the external force vanish in the thermodynamic limit, and $F(w)$ converges to a constant value. This implies that while avalanches induce local fluctuations in $F(w)$, the average global response becomes smooth for sufficiently large systems. This behavior is indeed observed in the standard depinning model in both $D = 1$ and $D = 2$: the force $F(w)$ remains approximately constant, and the amplitude of its fluctuations decreases with increasing system size (see black continuous lines in the lower panels of Fig. 1).

We now move to the case where relaxation is present ($\Theta \neq 0$), starting from $D = 1$. In this regime, the avalanche size distribution deviates significantly from that of standard depinning: it exhibits a broad, flat bump at large sizes and a steeper decay at small sizes (Fig. 1a), indicating a larger effective exponent $\tau$. The position of the bump, $S_{\max}$, remains controlled by the stiffness $k_0$, and does not depend on the system size $L$. This behavior is clearly illustrated in Fig. 2a, where we plot $S_{\max}$ as a function of $k_0$ for different system sizes. As $L$ increases, the data approach a clean power-law scaling $S_{\max} \sim k_0^{-\sigma}$, with an exponent $\sigma \simeq 2.05$, significantly larger than the value observed in the standard depinning case. Consistently, the global force $F(w)$ shows a nearly constant behavior with small fluctuations that diminish as the system size increases (thinner and thicker red curves in Fig. 1c). The average value of $F(w)$ lies between the values observed in the depinning model without relaxation, evaluated in two limiting cases: (i) the *soft interface*, corresponding to the post-relaxation configuration with reduced elastic constants $G_{ij}(1-\Theta)$; and (ii) the *rigid interface*, representing the pre-relaxation configuration with the original elastic constants $G_{ij}$.

We now turn to the 2D case with relaxation, where a novel dynamical scenario emerges. The avalanche size distribution $P(S)$ becomes bimodal for sufficiently small values of $k_0$. Specifically, after an initial power-law decay at small sizes—with an exponent $\tau$ larger than that of the standard depinning model—a secondary peak appears, corresponding to a cluster of very large system-spanning avalanches (highlighted by the ellipse in Fig. 1b). More precisely, we identify a critical stiffness value $k_c$, such that for $k_0 \geq k_c$, the distribution $P(S)$ exhibits a single power-law behavior with a finite-size cutoff $S_{\max}$ that diverges as $k_0 \to k_c^+$. Below this threshold ($k_0 < k_c$), the cutoff decreases again, but large-scale system size events start to be present, signaling the emergence of a distinct dynamical regime. This transition is also evident in the global force signal $F(w)$, since, for $k_0 < k_c$, $\langle \delta w \rangle$ is finite also in the thermodynamic limit, and $F(w)$ displays quasi-periodic oscillations between two well-defined values (Fig. 1d). These values are close to those observed in the depinning model with the soft and rigid interface, as defined before, and remain independent of system size. The divergence of $S_{\max}$ near $k_c$ is shown in Fig. 2b, confirming the presence of a finite critical point associated with the onset of a stick-slip-like regime. This behavior is reminiscent of what is observed in mean-field models, where a finite $k_c$ also marks a dynamical transition. However, in contrast to the fully connected case—where $k_c$ increases with $\Theta$, but the scaling exponents remain fixed at their depinning values ($\tau = 3/2$, $\sigma = 2$)—here both $\tau$ and $\sigma$ are significantly larger and *explicitly depend on* $\Theta$.

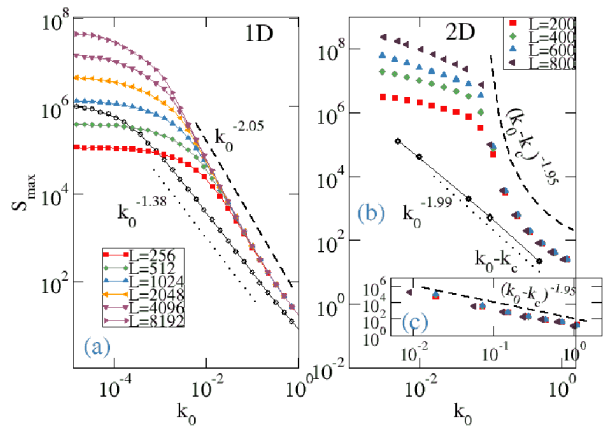

FIG. 2. $S_{\max}$ as a function of $k_0$. (a) 1D long-range. Black circles represents the depinning case ($\Theta = 0$) with $L = 8192$. Other symbols and colors are used for $\Theta = 0.5$ and different values of $L$, from bottom to top: $L = 2^8, \ldots, 2^{13}$. A power-law divergence is observed as $k_0 \to 0$. (b) The same of panel (a) for the 2D long-range model. Black circles for $\Theta = 0$ and different symbols and colors for $\Theta = 0.5$ with from bottom to top: $L = 200, 400, 600, 800$. The power law divergence occurs at $k_c = 0.11$. (c) Plot of same data of panel (b) vs $k_0 - k_c$.

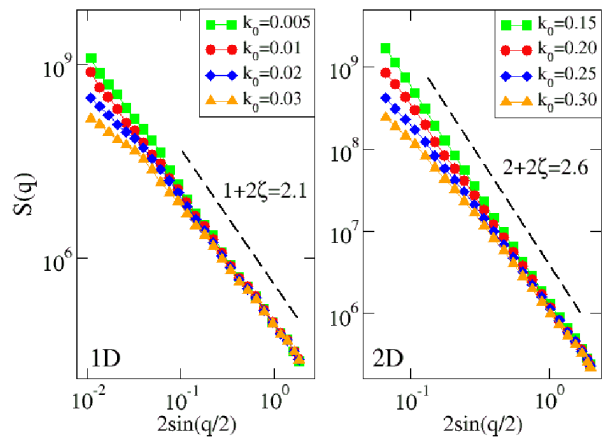

FIG. 3. Structure factor for 1D-Long range elasticity (left) and 2D-Long range elasticity (right). A consistent power appears when $k_0 \to 0$ (1D), or $k_0 \to k_c$ (2D). This allows to obtain the roughness exponent $\zeta$ as $\zeta \simeq 0.55$ in 1d, and $\zeta \simeq 0.3$ in 2D.

TABLE I. Critical exponents of the avalanche dominated regime. For 2D long-range elasticity we obtain $k_c = 0.11$.

| Exponent | $\alpha = 1$ & $D = 1$ | | $\alpha = 1$ & $D = 2$ | |
|---|---|---|---|---|
| | Depinning | $\Theta = 0.5$ | Depinning | $\Theta = 0.5$ |
| $\zeta$ | 0.39 | 0.55(5) | 0 | 0.30(1) |
| $\sigma$ | 1.38 | 2.05(5) | 2 | 1.95(5) |
| $\tau$ | 1.28 | - | 1.5 | 1.7(1) |

These results demonstrate that the presence of dynamical effects modifies the statistics of avalanches and alters the critical exponents. To complement this analysis, we characterize the shape of the elastic interface. In the stationary regime, disorder induces strong fluctuations that grow with distance as $h_r - h_0 \sim r^\zeta$, where $\zeta$ is the roughness exponent. A practical way to determine this exponent is through the structure factor,

$$S(q) \equiv \langle |\hat{h}_q|^2 \rangle \sim q^{-(D+2\zeta)},$$

as shown in Fig. 3, where $\hat{h}_q$ is the Fourier transform of the displacement field. According to the predictions of Ref. [15], the interface should become rougher as $\Theta$ increases. This is clearly observed in 2D, where the interface develops a positive and sizable roughness exponent $\zeta$. In contrast, in 1D the increase in roughness is less pronounced.

A summary of the measured critical exponents $\zeta$, $\sigma$, and $\tau$ is reported in Table I. In the standard depinning framework, the exponents $\tau$ and $\sigma$ are constrained by scaling relations involving the spatial dimension $d$, the roughness exponent $\zeta$, and the exponent $\alpha$ of the range of the interaction. In the anomalous regime, however, it remains unclear whether such relations persist or the exponents become independent.

The broad bump in the avalanche distribution $P(S)$ in the 1D long-range case deserves further investigation. We present a closer look at these anomalous avalanches in Fig. 4.

In the main panel, we compare the sequence of unstable blocks during two representative large avalanches: one from the standard depinning model and one from the model with relaxation. In the depinning case, the instability propagates as an anomalous diffusion process starting from a hypocenter and repeatedly revisiting the origin. In contrast, in the presence of relaxation, the instability quickly becomes ballistic, propagating with a well-defined velocity and concentrating on the active front of the avalanche. This difference in dynamics is also evident in the averaged behavior of avalanche propagation: in the depinning case, the average diameter grows superdiffusively as $d \sim t^{1/z}$ with a dynamic exponent $1 < z < 2$ [8, 10], while in the model with relaxation, the growth becomes ballistic.

The excess of large anomalous avalanches observed in the 1D model with relaxation appears as a broad bump in the distribution $P(S)$, and can be interpreted as a failed attempt of synchronization. Beyond a small nucleation size, the avalanche grows deterministically, similarly to the supercritical regime in 2D for $k_0 < k_c$. However, in 1D, as the unstable portion of the line advances, the increasing local elastic restoring force prevents the complete synchronization of the system. A similar behavior has been reported [40] near the cutoff of the avalanche distribution in standard depinning models [41]. We remark that our findings for 1D long range interaction models are qualitatively similar in 2D models with short range interaction (see Supplementary material, section II).

In summary, we have shown that non-overdamped dynamics in finite-dimensional systems give rise to two distinct scenarios. For short-range elasticity, or for long-range elasticity in one dimension, the system exhibits a pinned regime characterized by anomalous avalanches, followed by a flowing regime. In these cases, we do not expect hysteresis loops when the external force is controlled. In contrast, in the two-dimensional long-range case, an intermediate stick-slip regime emerges, marked by system-spanning reorganizations. Here, hysteresis loops are expected under a force-controlled protocol.

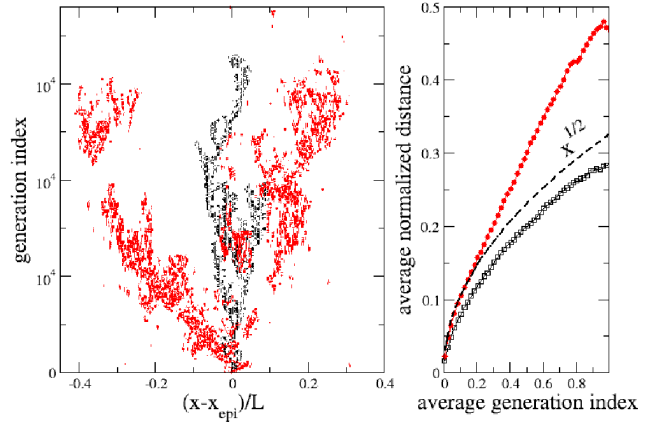

FIG. 4. (left panel) Avalanche profiles in 1D long-range elasticity. We show two representative avalanches with total size $S \simeq 20000$ for the cases $\Theta = 0$ (black open squares) and $\Theta = 0.5$ (red filled circles), simulated in a system of size $L = 1024$ and stiffness parameter $k_0 = 0.01$. Each point corresponds to a local instability within the avalanche, plotted with its position translated from the avalanche epicenter (horizontal axis) and its generation index (vertical axis). The horizontal axis is divided by the system size $L$. (Right panel) Analysis of $K = 10^6$ avalanches with sizes in the range $S \in [2 \times 10^5, 2.2 \times 10^5]$. For each avalanche and at each generation time, we compute the average displacement, defined as the absolute distance of toppling sites from the epicenter. These displacements are then averaged over all avalanches and plotted as a function of the normalized generation time (i.e., generation time divided by the total duration of each avalanche). The resulting average displacement is further normalized by the system size $L$. Black open circles correspond to the depinning case $\Theta = 0$, while red open squares refer to $\Theta = 0.5$. The dashed black line represents the reference diffusion scaling, $x^{1/2}$.

These results were obtained in the regime of slow relaxation, which tends to destabilize more sites than the fast relaxation studied in [16–18]. We therefore expect that synchronization remains absent under fast relaxation in short-range systems and in the 1D long-range case. For the 2D long-range case, further numerical simulations are needed; however, it is plausible that synchronization only occurs above a finite threshold $\Theta_c$.

Our study has focused on the depinning transition. However, yielding transitions are governed by a long-range, strongly anisotropic stress redistribution kernel.

In 2D and 3D, it is plausible that synchronization also occurs due to the long-range nature of interactions. Nevertheless, the anisotropy should localize instabilities into shear bands of dimension $D - 1$ rather than destabilizing the entire system. Such shear bands are indeed observed in materials under shear [12, 13, 25].

We acknowledge fruitful discussions with Vincenzo Maria Schimmenti, Adrien Rezzouk, and Francisco Pereyra Aponte. AR acknowledges funding by the ANR with reference ANR-23-CE30-0031-04.

# Supplementary Information
# Anomalous Critical Behavior of Driven Disordered Systems Beyond the Overdamped Limit

Giuseppe Petrillo,[1] Eduardo Jagla,[2] Eugenio Lippiello,[3] and Alberto Rosso[4]

[1]*Earth Observatory of Singapore (EOS), Nanyang Technological University, Singapore*
[2]*Centro Atomico Bariloche*
[3]*Department of Mathematics and Physics, University of Campania "L. Vanvitelli", Caserta, Italy*
[4]*LPTMS, CNRS, Université Paris-Saclay, Orsay, 91405, France*

## I. FURTHER DETAILS FOR THE LONG-RANGE CASE

To further validate our results, we include in this section additional measurements of the roughness exponent $\zeta$ for the long-range case. In particular, we provide results obtained by analyzing the fractal geometry of individual avalanches.

In the main text, $\zeta$ was extracted from the power spectrum $S(q)$. Here, we use an alternative method based on avalanche shapes. For each avalanche, we define:

- The size $S$, i.e., the total number of instabilities (already introduced in the main text);

- The spatial extension $A$, defined as the number of distinct sites that were unstable at least once;

- A characteristic length $\xi$, defined via $\xi = A^{1/D}$.

Assuming scale-invariance, we expect the scaling relation:

$$S \sim \xi^{D+\zeta}, \tag{1}$$

which provides an independent estimate of $\zeta$ as shown in the two panels of Fig. S1.

## II. THE SHORT-RANGE ELASTICITY

In this section, we present the same results already shown for the long-range case, but now for the short-range case ($\alpha = 2$), which can also be studied using the discrete Laplacian: $G_{ij} = 1/D$ if $i$ and $j$ are nearest neighbors, and zero otherwise. Our results are qualitatively similar to the 1D long-range case: the avalanche size distribution develops a broad bump (Fig. S2), and $S_{max}$ diverges only when $k_0 \to 0$ (Fig. S3). The roughness of short-range interfaces is shown in Fig. S4 and Fig. S5. As in the long-range case, the roughness exponents are slightly larger than at depinning. We also observe that, in the 1D short-range case, the interface appears flatter at large scales compared to both the 2D case and the 1D long-range case. As in the long-range case, the bump in the roughness is caused by the presence of avalanches expanding ballistically (see Fig.S6).

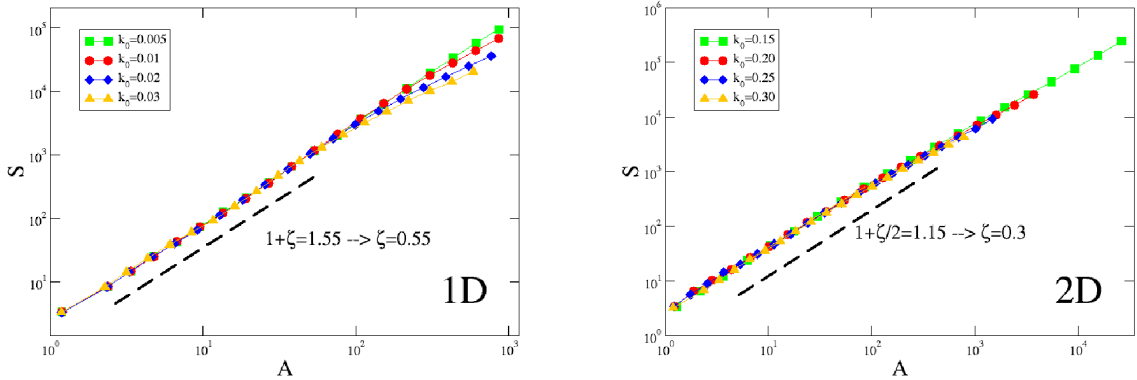

FIG. S1. Verification of the roughness exponent at $\Theta = 0.5$. **Left:** 1D case with system size $L = 1024$, using $10^6$ avalanches. The roughness exponent obtained in the main text, $\zeta = 0.55 \pm 0.1$, is consistent with the scaling law of Eq. 1. **Right:** 2D case with system size $L = 200$, based on $10^6$ avalanches. The exponent $\zeta = 0.30 \pm 0.01$ from the main text again matches the scaling behavior observed here.

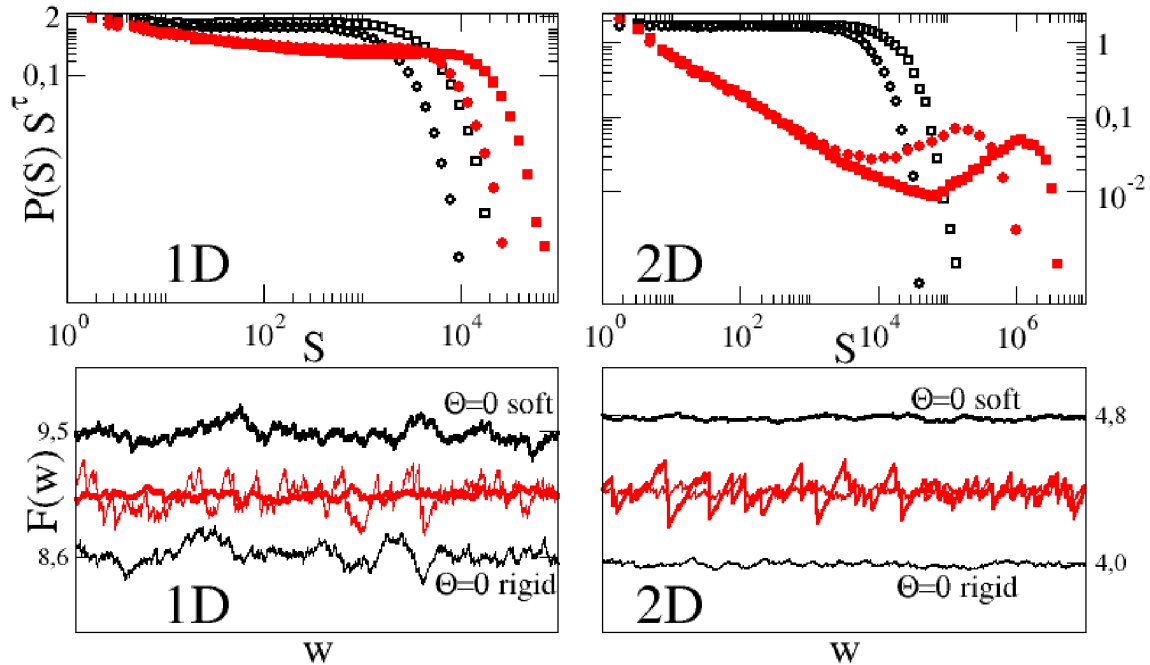

FIG. S2. Equivalent of Fig. 1 from the main text, for the case of short-range elasticity. We use $L = 1024$ in 1D and $L = 400$ in 2D with $k_0 = 0.01$ and $\Theta = 0$ (open black circles), $k_0 = 0.005$ and $\Theta = 0$ (open black squares), $k_0 = 0.01$ and $\Theta = 0.5$ (filled red circles) and $k_0 = 0.005$ and $\Theta = 0.5$ (filled red squares). Note that in the 2D case, smaller values of $k_0$ are used compared to the long-range elasticity scenario, in order to obtain comparable sizes for the largest avalanches. (Inset) $F(w)$ for $K_0 = 0.01$ and $\Theta = 0.5$, is plotted in red with thinner line for $L = 1024$ (1D) and $L = 400$ (2D) and thicker line for $L = 8192$ (1D) and $L = 800$ (2D). Black line are for the depinning models ($\Theta = 0$) for the rigid $k_0 = 0.01$ (lower curve) and soft $k_0 = 0.005$ (upper curve) interface.

We now characterize the geometry of large avalanches. As shown in Fig. S6, even in the short-range case, rare large events originating from a flat initial profile (flat-bump protocol) display a ballistic expansion. The left panel shows an example of such an avalanche. The right panel confirms that the roughness exponent $\zeta$ extracted from the power spectrum $S(q)$ remains close to the depinning value.

A summary of the measured critical exponents $\zeta$, $\sigma$, and $\tau$ is reported in Table I. In the standard depinning framework, these exponents are connected by scaling relations:

$$\tau = 2 - \frac{\alpha}{d+\zeta}, \qquad \sigma = \frac{\alpha}{d+\zeta} \cdot \alpha. \tag{2}$$

In our protocol with finite $\Theta$, the scaling relations are no longer valid, yet the deviations are moderate, especially for $\zeta$.

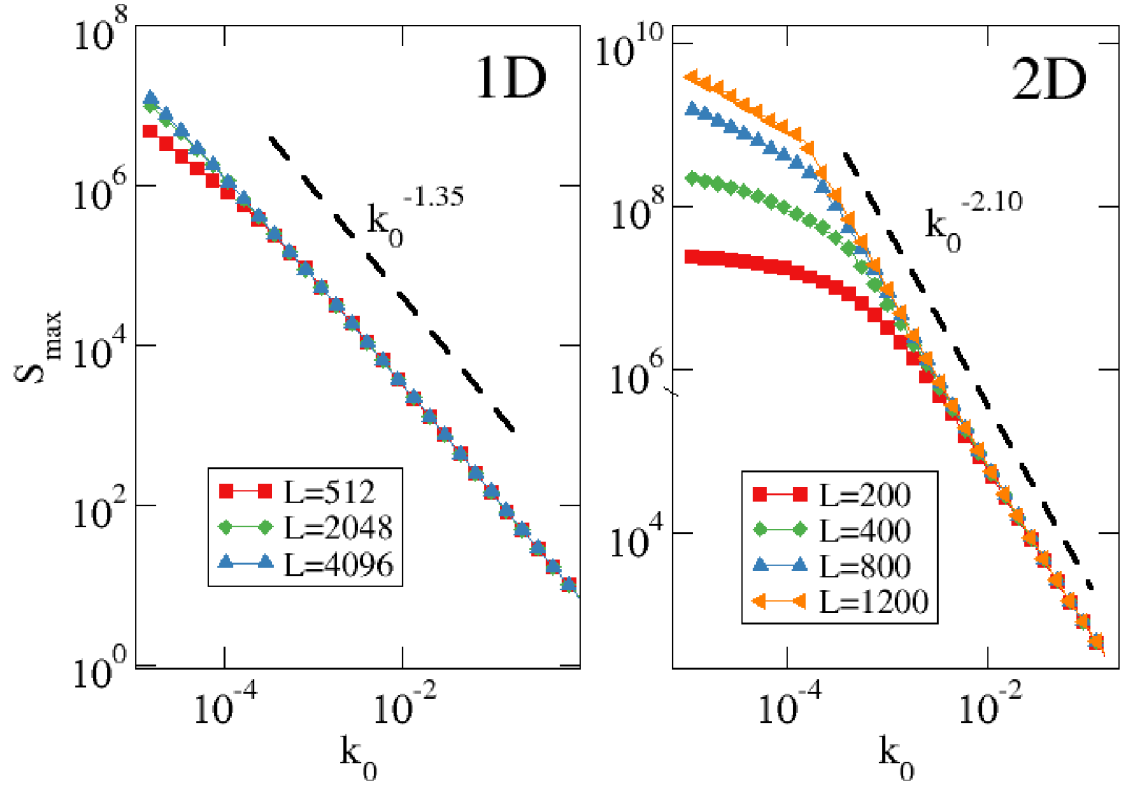

FIG. S3. Equivalent of Fig. 2 from the main text representing the largest avalanche size $S_{\max}$ on the kick amplitude $k_0$ at fixed $\Theta = 0.5$ for short-range elasticity.

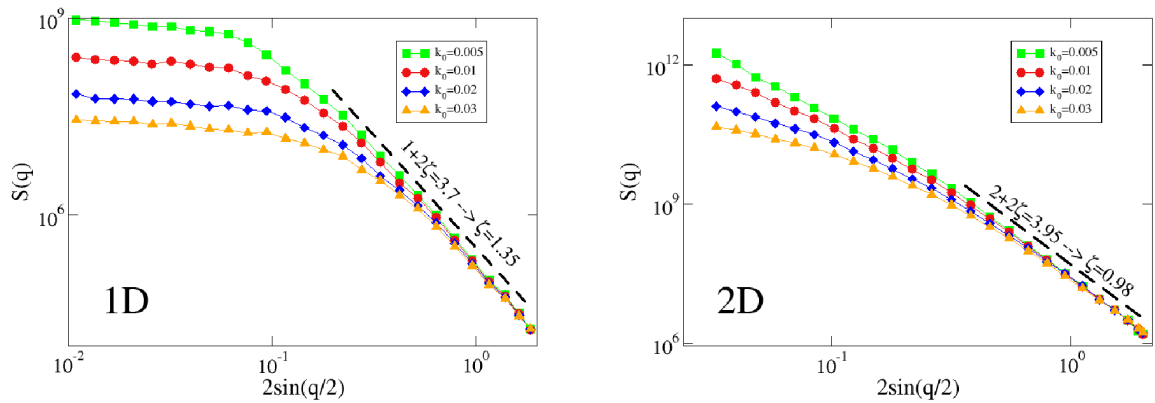

FIG. S4. Equivalent of Fig. 3 from the main text, now shown for the case of short-range elasticity. Left panel is for the 1D case ($L = 1024$) and right panel for the 2D case ($L = 400$). Note that in the 2D case, smaller values of $k_0$ are used compared to the long-range elasticity scenario, in order to obtain comparable sizes for the largest avalanches.

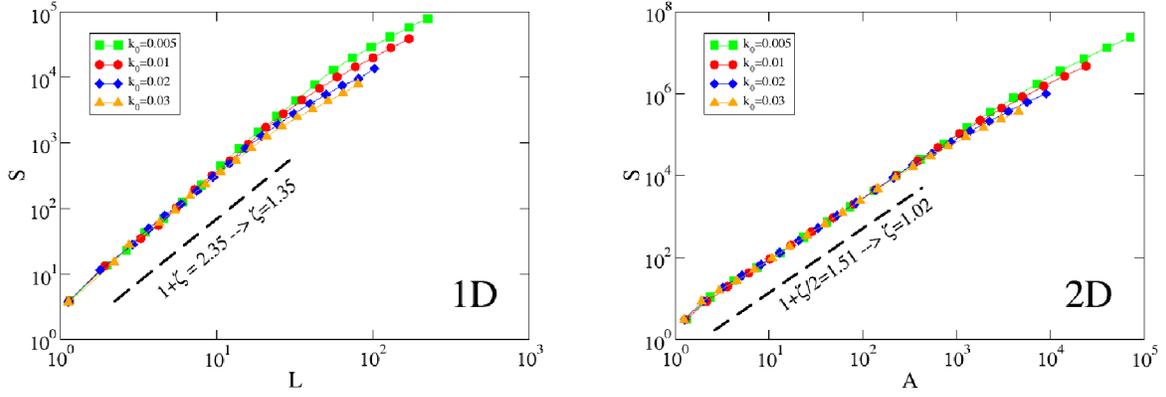

FIG. S5. Equivalent of Supp. Fig. 1, now shown for the case of short-range elasticity. Left panel is for the 1D case ($L = 1024$) and right panel for the 2D case ($L = 400$). In the 2D case, smaller values of $k_0$ were chosen compared to the long-range case, in order to obtain comparable sizes for the largest avalanches.

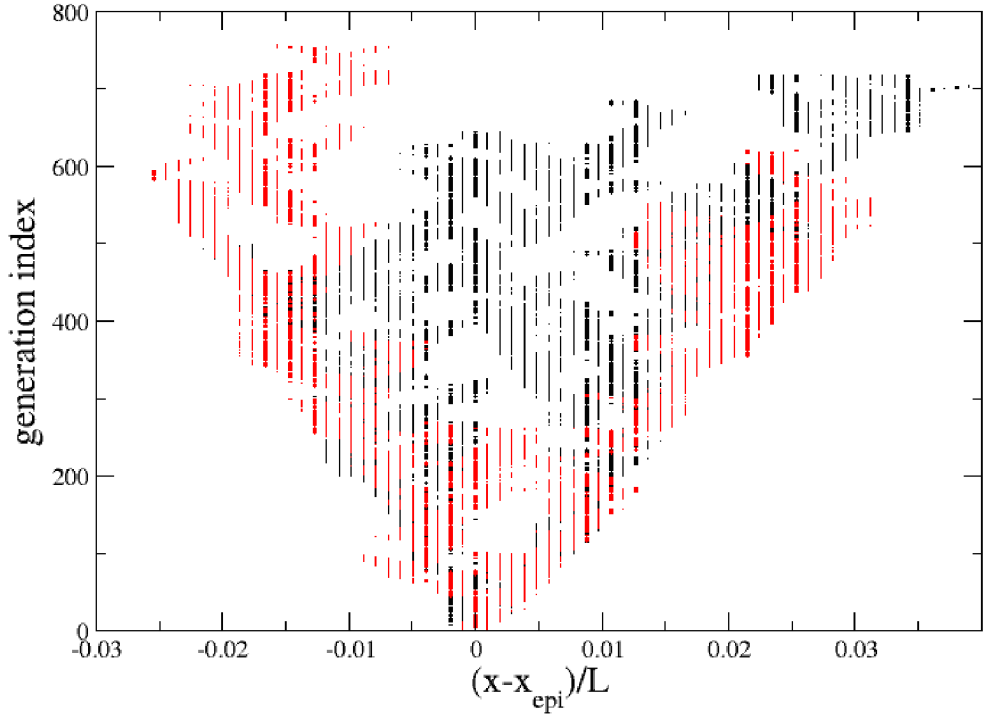

FIG. S6. The equivalent of the left panel of Fig.4 of the main text for the avalanche profiles in 1D for short-range elasticity. We show two representative avalanches with total size $S \simeq 20000$ for the cases $\Theta = 0$ (black open squares) and $\Theta = 0.5$ (red filled circles), simulated in a system of size $L = 1024$ and stiffness parameter $k_0 = 0.01$.

TABLE I. Critical exponents for short-range elasticity ($\alpha = 2$).

| Exponent | $D = 1$ | | $D = 2$ | |
|:---:|:---:|:---:|:---:|:---:|
| | Depinning | $\Theta = 0.5$ | Depinning | $\Theta = 0.5$ |
| $\zeta$ | 1.25 | 1.35 (1) | 0.752 | 1.00(2) |
| $\sigma$ | 1.125 | 1.35(1) | 1.38 | 2.10(2) |
| $\tau$ | 1.11 | 1.56(1) | 1.27 | 1.70(2) |